# Intrinsic anomalous Hall effect in Ni-substituted magnetic Weyl semimetal $Co_3Sn_2S_2$

Gohil S. Thakur[1], Praveen Vir[1], Satya N. Guin[1], Chandra Shekhar[1], Richard Weihrich[2], Yan Sun[1], Nitesh Kumar[1*] and Claudia Felser[1*]

[1]Max-Planck-Institute für Chemische Physik Fester Stoffe, 01187 Dresden, Germany

[2]Universität Augsburg, IMRM, Universitätsstraße 2, 86135 Augsburg, Germany

**ABSTRACT:** Topological materials have recently attracted considerable attention among materials scientists as their properties are predicted to be protected against perturbations such as lattice distortion and chemical substitution. However, any experimental proof of such robustness is still lacking. In this study, we experimentally demonstrate that the topological properties of the ferromagnetic kagomé compound $Co_3Sn_2S_2$ are preserved upon Ni substitution. We systematically vary the Ni content in $Co_3Sn_2S_2$ single crystals and study their magnetic and anomalous transport properties. For the intermediate Ni substitution, we observe a remarkable increase in the coercive field while still maintaining significant anomalous Hall conductivity. The large anomalous Hall conductivity of these compounds is intrinsic, consistent with first-principle calculations, which proves its topological origin. Our results can guide further studies on the chemical tuning of topological materials for better understanding.

## 1. INTRODUCTION

Topological materials are a new class of quantum materials, whose surface electronic states are protected against weak structural perturbations.[1-4] For example, the two-dimensional graphene-like Dirac cones at the surface of topological insulators are protected until the bulk electronic structure is destroyed.[1,2] The surface states of the Weyl semimetals are more exotic as they appear in the form of arcs, which can be destroyed only if the Weyl points having opposite chiralities in the bulk are brought together and annihilated by means of large structural deformations.[5,6] Recently, several Weyl semimetals have been theoretically predicted and experimentally confirmed using spectroscopic and electrical transport techniques.[7] Most of these compounds are nonmagnetic, where the non-centrosymmetric crystal structure is an essential requirement to attain the Weyl points. However, the Weyl points can also exist in centrosymmetric systems if the compound is magnetic. The Weyl semimetals have recently garnered enormous interest owing to their exotic transport properties.[8-10] Nonmagnetic Weyl semimetals exhibit large mobility of the charge carriers and extremely high magnetoresistance[8,9] whereas a giant anomalous Hall effect is observed in magnetic Weyl semimetals.[11,12] The large room-temperature mobility of $MPn$ (M = Ta and Nb; $Pn$ = P and As) family of Weyl semimetals is responsible for their excellent hydrogen evolution catalytic activities.[13]

The layered Shandite compound $Co_3Sn_2S_2$ is one of the very few known magnetic Weyl semimetals.[14-16] Apart from its interesting topological properties, this compound is also investigated for its thermoelectric properties and possible skyrmionic phase.[17] The topological effect evoked by the Weyl points is responsible for the giant anomalous Hall effect in this compound.[11,12] Although the transport properties of several Weyl semimetals have been recently investigated, the behavior of the Weyl points upon chemical substitution has not been studied, mainly due to the unavailability of single crystals of the substituted compositions. In this study, we investigate the effects of Ni-substitution on the electrical transport properties of the magnetic Weyl semimetal $Co_3Sn_2S_2$. We find that although the magnetization of $Co_{3-x}Ni_xSn_2S_2$ decreases with increasing Ni content, the total anomalous Hall conductivity remains intrinsic to the band structure. This emphasizes the fact that the topological effects in $Co_{3-x}Ni_xSn_2S_2$ are robust against varied Ni-substitution.

## 2. EXPERIMENTS AND PROCEDURE

$Co_3Sn_2S_2$ melts congruently at 1163 K and hence single crystals were grown by the solidification of the melt upon slow cooling. Single crystals with the nominal compositions of $Co_{3-x}Ni_xSn_2S_2$ ($x$ = 0 to 0.6) were synthesized by using the solid-state sealed-tube method. Stoichiometric amounts of the elements (~10 g) were weighed inside an Ar-filled glove box ($H_2O$ and $O_2$ < 0.1 ppm) and placed in alumina crucibles, which were subsequently sealed in quartz tubes under a ⅓rd argon atmosphere. The tubes were placed vertically and heated in a programmable muffle furnace to 673 K in 8 h with a dwell time of 10 h, and then the temperature was increased to 1323 K in 4 h. The components were allowed to melt and homogenize for 24 h. Subsequently, the tubes were slowly cooled to 1073 K in 72 h, and then naturally cooled by switching off the furnace. The product appeared as a shiny silvery ingot. Many black plates with mirror surfaces could be easily extracted from the ingot by mechanical cleaving (Fig. S1 in Supplementary Information). The flat surfaces of the crystals correspond to the *ab*–plane, as adjudged by Laue diffraction patterns, which is common for many such layered materials (Fig. S2 in Supplementary information). The phase purity and crystallinity of each sample were evaluated by powder X-ray diffraction. The crystals were polished and cut to thin rectangular bars for transport measurements. Energy dispersive spectroscopy and chemical analysis were employed to determine the exact compositions of all the samples, which were very close to the nominal compositions (Tables S1 and S2 in Supplementary information). The magnetization of single crystals was measured in applied magnetic fields up to $\mu_o H = 7$ T and in the temperature range between 2 and 300 K using a SQUID MPMS-3 magnetometer (Quantum Design). Temperature-dependent resistivity was measured on single crystals in a PPMS instrument (Quantum Design) in the temperature range of 2 to 300 K under an applied field up to $\mu_o H = 9$ T.

The first principle calculations were performed using the Vienna Ab initio Simulation Package (VASP) with the projected augmented wave method.[18] The exchange and correlation energies were considered in the generalized gradient approximation (GGA) level.[19] The atoms Ni were put on the original Co-sites, with the lattice structures and magnetic moments relaxed. After relaxation, we chose the final lattice and magnetic structure with the minimum total energies for each composition. For each composition, the Bloch wave-functions were projected onto maximally localized Wannier functions starting with the atomic Co-*s*, Co-*d*, Sn-*p*, and S-*p* orbitals. The Tight-binding model Hamiltonians were then constructed accordingly.[20] By using the tight-binding model Hamiltonian, the intrinsic AHC was computed in the linear response Kubo formula approach.[21]

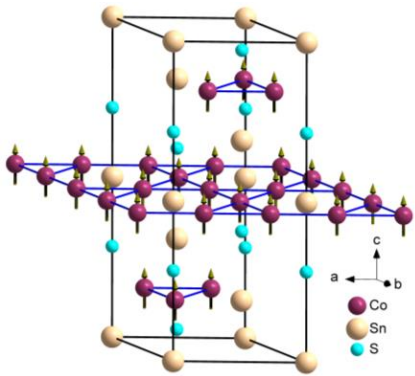

Figure 1. Crystal structure of $Co_3Sn_2S_2$ with a kagomé layer of magnetic cobalt atoms (purple spheres); the cyan and beige spheres represent S and Sn atoms, respectively.

3. EXPERIMENTAL RESULTS AND ANALYSIS

3.1 **Structure**. $Co_3Sn_2S_2$ crystallizes in the rhombohedral system in the $R\bar{3}m$ space group common for all the shandites.[22] Fig. 1 shows the crystal structure of $Co_3Sn_2S_2$. Co atoms form a kagomé layer within the Co–Sn layers. Ni substitution does not change the structure as the other end member $Ni_3Sn_2S_2$ also crystallizes in the same shandite structure and hence the formation of a solid solution $Co_{3-x}Ni_xSn_2S_2$ in the entire range ($x = 0$ to 3) is possible.[22] The changes in lattice parameters are expectedly not large as the difference in ionic radius between Co and Ni is very small, as well as the doping level.[23] However the *a*-parameter increases steadily (from 5.366 Å for $x = 0$ to 5.373 Å for $x = 0.6$) which leads to volume expansion upon Ni substitution. The $d_{Co\text{-}Co}$ increases consequently.

3.2 **Magnetic Properties**. $Co_3Sn_2S_2$ is a ferromagnet with a Curie temperature ($T_C$) of ~180 K and the *c*-axis as the easy axis of magnetization.[24,25] Fig. 2(a) shows the magnetization of the pristine and Ni-substituted compounds as a function of the temperature at a fixed magnetic field of 0.1 T. The sharp increase in the magnetization of $Co_3Sn_2S_2$ below 178 K ($T_C$) corresponds to the ferromagnetic ordering, which agrees well with the earlier reports.[24-26] $T_C$ shifts toward lower temperatures upon increasing Ni content, while magnetization at low temperature decreases, consistent with the results for polycrystalline samples.[26] All samples exhibit clear ferromagnetic transitions except that with $x = 0.6$, where the ferromagnetic interactions are almost suppressed. The inset of Fig. 2(a) shows an almost linear decrease in $T_C$ with the increase in Ni content. Fig. 2(b) shows the magnetization as a function of the magnetic field along the *c*-axis at 2 K. A saturated magnetic moment ($M_s$) of ~0.9 $\mu_B$/f.u. and coercivity ($H_c$) of 0.31 T is observed for the pristine $Co_3Sn_2S_2$. The large coercivity indicates a hard and anisotropic ferromagnetic character of the compound. Upon Ni substitution, the saturated moment decreases to the lowest value of ~0.03 $\mu_B$/f.u. at $x = 0.6$ (Fig. 2(c)). Ni−substitution decreases magnetism through following multiple effects; 1) upon Ni−substitution the distance between the nearest Co atoms ($d_{Co\text{-}Co}$) of the kagomé lattice increases which weakens the magnetic coupling. 2) Since the valence state of Ni in $Ni_3Sn_2S_2$ is $Ni^0$, it has fully filled *d*-bands which means it is non-magnetic.[27] Substituting a non-magnetic substituent at the Co sites in $Co_3Sn_2S_2$ breaks the magnetic coupling between the nearest Co atoms thereby suppressing the magnetism. Also, in $Co_3Sn_2S_2$, the energy gap in exchange-split 3*d* states of Co atoms is at a slightly lower binding energy (BE) as compared to the 3*d* states of Ni in $Ni_3Sn_2S_2$.[28] Thus, the transition to ferromagnetism is suppressed upon Ni-substitution. The coercivity first increases gradually for compositions with $x$ up to 0.2, and then suddenly rises to the maximum of ~1.2 T at $x = 0.4$ and 0.45, and then is considerably decreased at $x = 0.6$. The increase in the coercivity of the single-crystalline Ni-substituted sample is probably attributed to the pinning of the spins by the Ni inclusions in the kagomé lattice. At higher Ni content (> 0.45) the magnetism becomes vanishingly small reducing the $H_c$ significantly and hence giving rise to a peak-like behavior in $H_c$ vs $x$ curve.

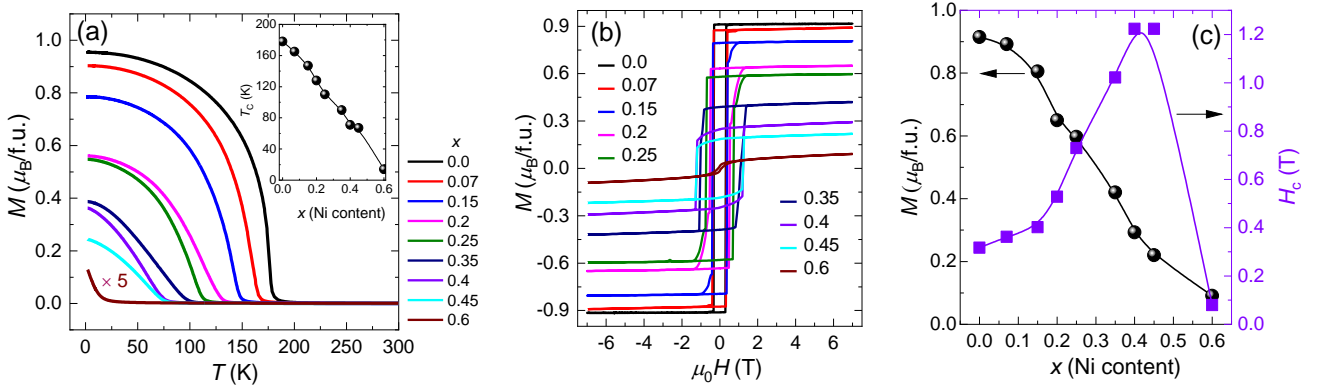

Figure 2. Magnetization curves for $Co_{3-x}Ni_xSn_2S_2$ ($x$ = 0 to 0.6). (a) Temperature-dependent magnetization; the inset shows the decrease in $T_C$ with the increase in Ni content, (b) field-dependent magnetization loops, and (c) variations in the magnetic moment and coercive field ($H_c$) with the Ni content.

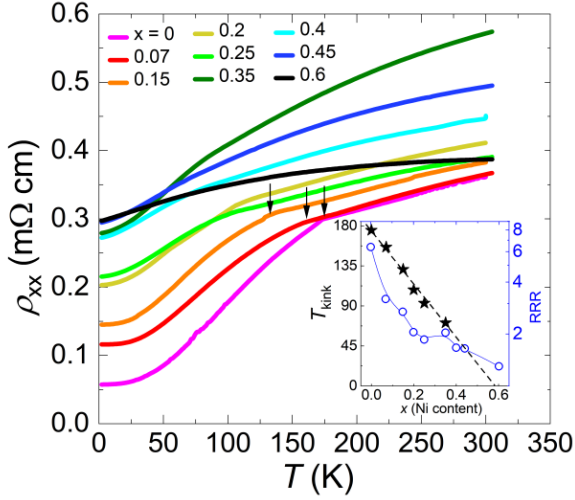

Figure 3. Temperature-dependent resistivity ($\rho_{xx}$) of all $Co_{3-x}Ni_xSn_2S_2$ compositions. The inset shows the change in the kink temperature and RRR with Ni content.

**3.3 Transport Properties.** The resistivity of all the samples was measured by applying a current in the $ab$-plane. All Ni-substituted compositions exhibit metallic behavior similar to that of the pristine $Co_3Sn_2S_2$, as shown in Fig. 3. The rate of change in resistivity is increased below $T_C$, which yields a kink-like feature at $T_C$. A trend similar to that of the magnetization is observed for the zero-field longitudinal resistivity ($\rho_{xx}$), where the kink shifts toward lower temperatures and smoothens at higher Ni concentrations (inset of Fig. 3). This kink practically remains unchanged upon the application of magnetic field (Figure S3 in the Supplementary information). At $x > 0.35$, the kink is completely smeared out and is not discernible at all. Although, no systematic trend of the variation in room-temperature resistivity ($\rho_{xx}^{300\,K}$) or residual resistivity ($\rho_{xx}^{2\,K}$) is observed, the residual resistivity ratio ($\rho^{300\,K}/\rho^{2K}$) decreases gradually with Ni content reflecting enhanced disorder induced by Ni-substitution (inset of Fig. 3). A steady decrease in the Hall constant and increase in charge carrier concentration is also observed as a consequence of Ni-substitution as the number of electrons in the system increase (Figure S4 in the Supplementary information).

In the Hall resistivity measurements, the current was passed along the $a$-axis, the magnetic field was applied along the $c$-axis, and the Hall voltage was measured perpendicular to the $a$-axis in the $ab$-plane. The Hall resistivity ($\rho_{xy}$) of each Ni-substituted sample as a function of the magnetic field exhibits a rectangular hysteresis loop similar to that of the magnetization at 2 K, as for the parent sample (see Fig. S5 in Supplementary information).

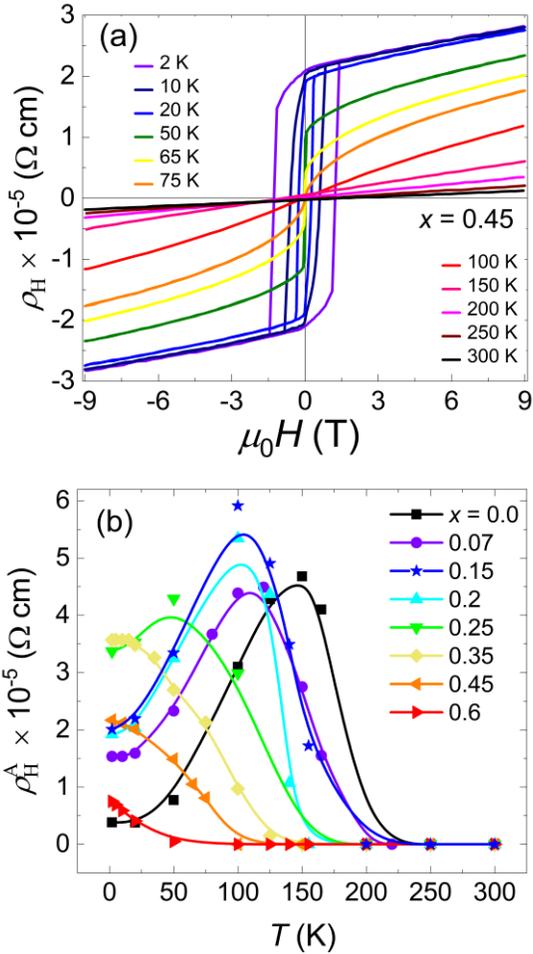

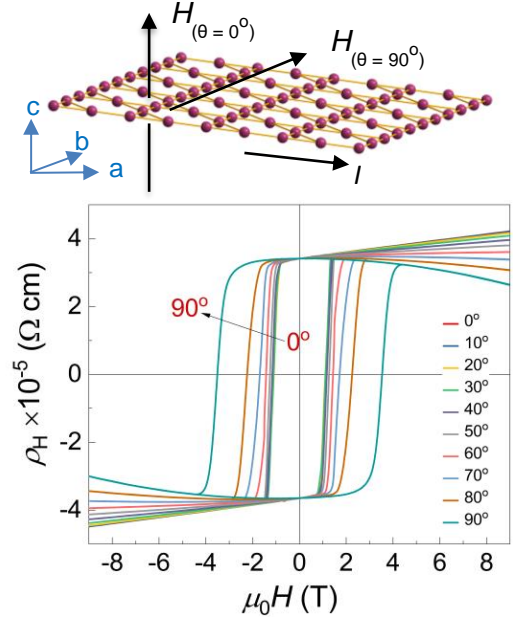

Figure 5. Angular dependence of the Hall resistivity of $Co_{3-x}Ni_xSn_2S_2$ ($x$ = 0.25). The upper inset shows the direction of the magnetic field and current in the crystal lattice.

Figure 4. (a) Variation in Hall resistivity of $Co_{3-x}Ni_xSn_2S_2$ with the magnetic field at $x$ = 0.45 and (b) temperature dependent anomalous Hall resistivity for various Ni concentrations.

Fig. 4(a) shows $\rho_H$ of the sample with $x$ = 0.45 at different temperatures. $\rho_H$ at zero magnetic field is defined as anomalous Hall resistivity ($\rho_H^A$). The coercivity of the hysteresis loop decreases with the increase in temperature with a loss of hysteresis in the vicinity of $T_C$. Above $T_C$, no anomalous Hall effect is observed and $\rho_H$ behaves linearly with the applied magnetic field without any hysteresis. Fig. 4(b) shows the temperature-dependent $\rho_H^A$ of samples with various Ni concentrations. For the pristine and low Ni-substituted compositions ($x$ < 0.25), $\rho_H^A$ starts increasing suddenly below $T_C$ up to the maximum, and then decreases down to 2 K. However, at $x$ = 0.35, no maximum is achieved below $T_C$ and $\rho_H^A$ continues to increase and then tends to saturate at the lowest temperature. Notably, no significant variations in maximum $\rho_H^A$ are observed in the $x$ range of 0 to 0.25. At $x$ = 0.6, $\rho_H^A$ continues to increase without saturation below $T_C$ (~20 K).

Fig. 5 shows the angle-dependent $\rho_H$ at 2 K for the sample with $x$ = 0.25. The current was passed along the $a$-axis, while the Hall voltage was measured perpendicular to the $a$-axis in the $ab$-plane. In the case of $\theta$ = 0°, the magnetic field was applied along the $c$-axis, which was perpendicular to both applied current and Hall voltage leads. At larger angles, the magnetic field was rotated towards the in-plane direction along the Hall voltage leads. The $\theta$ = 90° data correspond to the case where the magnetic field is directly along the direction of the Hall voltage leads. $\rho_H^A$ remains constant despite the variation in angle, while the coercive field increases with the angle and reaches a giant value of 3.5 T at $\theta$ = 90°. The constant value of $\rho_H^A$ is a clear indication of the out-of-plane magnetization in the compound.

The anomalous Hall conductivity ($\sigma_H^A$), defined as the finite $\sigma_H$ at zero magnetic field, is a more useful quantity to understand the anomalous transport properties of a compound, as its magnitude can be directly related to the band structure. As the conductivity is a tensor quantity, the Hall conductivity $\sigma_H$ is calculated as $\sigma_H = \frac{\rho_H}{\rho_H^2 + \rho^2}$. Fig. 6(a) shows the square hysteretic loops observed for $\sigma_H$ at various concentrations of Ni at 2 K. Fig. 6(b) shows the temperature-dependent $\sigma_H^A$ values of pristine and Ni-substituted samples. A marked contrast in the temperature dependence of $\sigma_H^A$ is observed, compared to the corresponding data for $\rho_H^A$ (Fig. 4(b)). Upon the decrease in temperature, $\sigma_H^A$ increases suddenly below $T_C$ and tend to saturate with the further reduction in temperature down to 2 K, instead of exhibiting a peak, as in the case of the temperature-dependent $\rho_H^A$. The corresponding $\sigma_H^A$ and anomalous Hall coercivity ($H_c^A$) are plotted in Fig. 6(c). The highest value of $\sigma_H^A$ = 1164 $\Omega^{-1}cm^{-1}$ is observed for the pristine $Co_3Sn_2S_2$, which decreases linearly upon Ni–substitution to the smallest value of $\sigma_H^A = 85\ \Omega^{-1}\text{cm}^{-1}$ for the nearly non-magnetic sample with $x = 0.6$. In contrast to $\sigma_H^A$, $H_c^A$ increases with the Ni content. Giant coercivities of ~1–1.2 T are observed at $x = 0.35$–$0.45$ together with remarkable $\sigma_H^A$ values in the range of 300–500 $\Omega^{-1}\text{cm}^{-1}$.

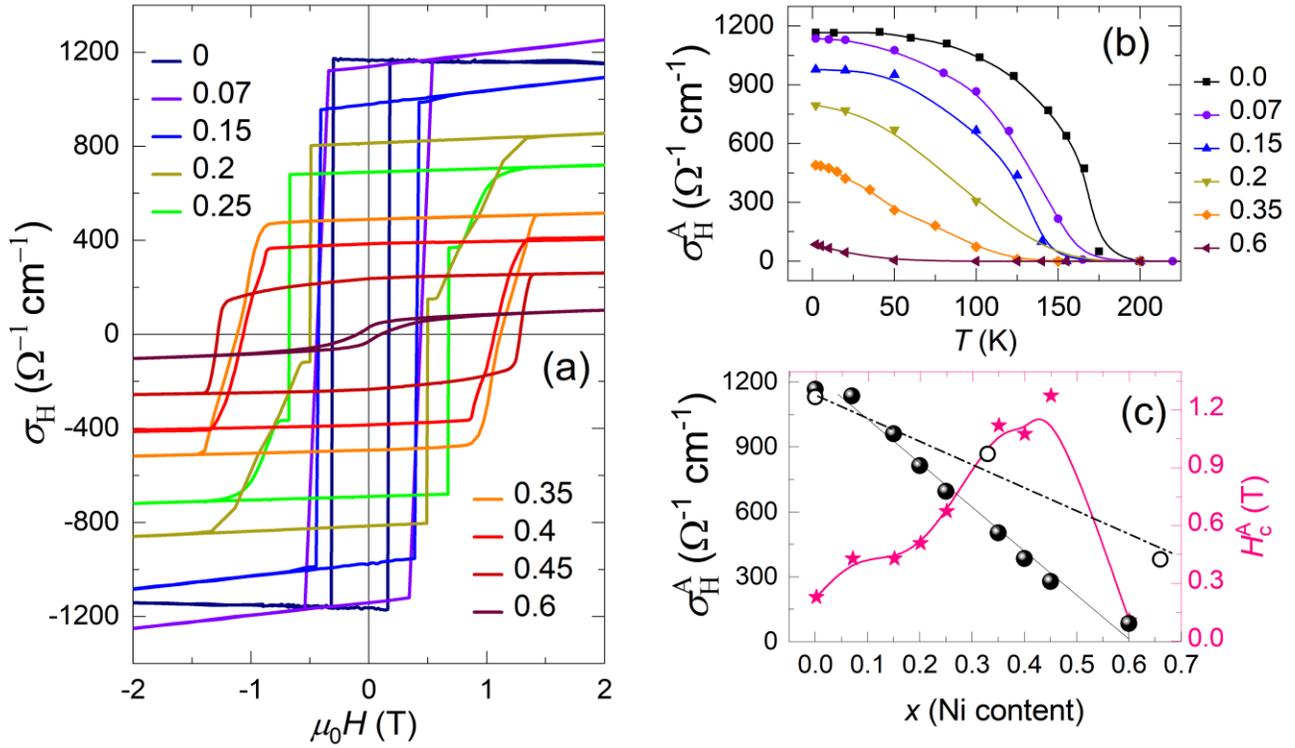

Figure 6. (a) Field-dependence of the Hall conductivity, (b) temperature dependence of the anomalous Hall conductivity (c) variation in anomalous Hall conductivity and coercivity with the Ni content in all $Co_{3-x}Ni_xSn_2S_2$ compositions; open circles are the $\sigma_H^A$ data point obtained from the band structure calculations.

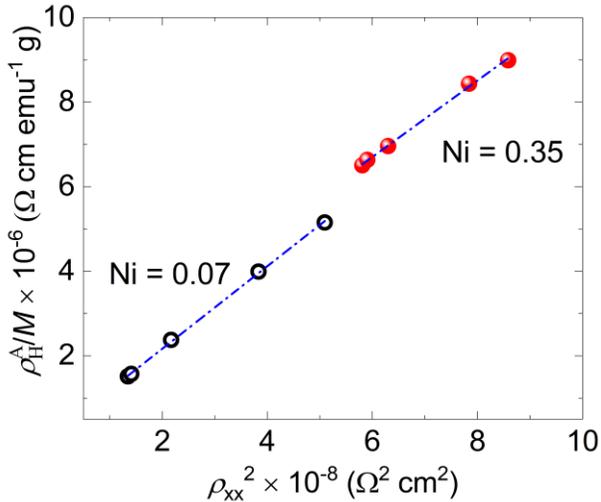

Figure 7. $\rho_H^A/M$ against $\rho^2$ for two compositions of $Co_{3-x}Ni_xSn_2S_2$ with $x = 0.07$ and $0.35$

## 4. DISCUSSION AND CONCLUSIONS

In order to demonstrate that the anomalous Hall conductivity of the Ni-substituted samples still originate from topological effects and not from impurity scattering processes, we analyze the intrinsic nature of the anomalous Hall conductivity. The almost constant $\sigma_{xy}$ (below $T_C$) with respect to $\sigma_{xx}$ and actual values of $\sigma_{xx}$ falling in the moderately dirty regime (3000–20000 $\Omega^{-1}\text{cm}^{-1}$) (see Fig. S6 in Supplementary information) strongly indicate the intrinsic nature of the anomalous Hall effect in all Ni-doped samples.[29] Furthermore, the intrinsic band-structure-originated anomalous Hall resistivity of a ferromagnet is directly proportional to the magnetization times the square of the longitudinal resistivity, i.e., $\rho_H^A \propto M\rho^2$. Fig. 7 shows $\rho_H^A/M$ against $\rho^2$ for two of the intermediate Ni-substituted compositions. The linear relation between these quantities further confirms the intrinsic topological origin of the anomalous Hall effect in the Ni-substituted $Co_3Sn_2S_2$. Additional details on the various contributions to the anomalous Hall effect are discussed in the Supplementary Information.

Ab-initio band-structure calculations were carried out to provide strong theoretical support to our results discussed above. Three cases of $Co_{3-x}Ni_xSn_2S_2$ were considered, magnetic ($x = 0$ and $0.33$), borderline magnetic ($x = 0.66$), and purely nonmagnetic ($x = 1.0$). The results are presented in figure S8 of the supplementary information. A peak of the energy-dependent $\sigma_H^A$ is observed almost at the Fermi surface for the pristine sample. With the increase in Ni content, this peak shifts farther from the Fermi surface (Fig. 8). The calculated magnetic moments decrease with the increase in Ni content, which is qualitatively consistent with the experimental measurements. The decrease in total magnetic moment directly leads to a decrease in intrinsic anomalous Hall conductivity (Fig. S7 in Supplementary information). For the magnetic samples ($x = 0$ to $0.66$), significant values of $\sigma_H^A$ are observed, whereas that of the nonmagnetic sample is zero. The decrease in $\sigma_H^A$ is attributed to the up-shift of the chemical potential upon the Ni doping.

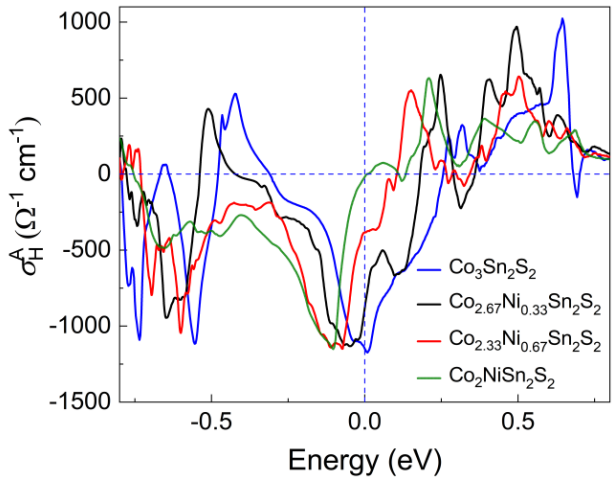

Figure 8. Energy dependence of $\sigma_H^A$ of $Co_{3-x}Ni_xSn_2S_2$ for $x = 0.0, 0.33, 0.66$, and $1.0$ obtained by a density functional theory calculation. The values of $\sigma_H^A$ at 0 eV (Fermi energy) are consistent with the experimental results.

For the pristine $Co_3Sn_2S_2$, the intrinsic AHE is mainly dominated by the band anti-crossings from topological band inversion. Owing to the mirror symmetry, gapless nodal lines form around the fermi level without considering spin-orbit coupling (SOC). These nodal lines are broken by SOC with opening band gaps (anti-crossing), meanwhile generating strong Berry curvatures around the nodal lines band anti-crossings. Upon Ni-substitution, the mirror symmetry is broken, and the nodal line linear band crossings vanish. However, the band inversion is robust as long as the band order doesn´t change. These band inversion with SOC also generate strong Berry curvature around the k-points with band anti-crossing (Fig S8 in SI). According to our calculations, the band inversion is maintained in the substitution range $x = 0$ to $1.0$. Therefore, the distribution of Berry curvature remains nearly unchanged in the reciprocal space. Since the band anti-crossings have dispersion in energy space, only the manner in which Fermi level cuts the anti-crossings change due to the band filling on Ni-substitution.

From our calculations, the effective cutting between Fermi level and nodal-line like band anti-crossing decreases as the Ni content increases. Furthermore, the magnitude of local Berry curvature also depends on the magnitude of the magnetic moments. The small magnetic moments from Ni also decrease the local Berry curvature. Therefore, the overall effect of Ni-substitution is to decrease both the local Berry curvature strength and the effective cutting with the Fermi level, leading to a decrease of intrinsic AHE. Hence, our calculations qualitatively agree well with the experimental results (see Fig. 6(c)).

In conclusion, we demonstrated that the Ni substitution in the ferromagnetic Weyl semimetal $Co_{3-x}Ni_xSn_2S_2$ had multiple effects on magnetic and transport properties. Although the magnetic moment continuously decreased with the substitution, the coercive field significantly increased. The effect of the large loop opening in the magnetization was also reflected in the Hall resistivity and Hall conductivity. The intermediate Ni-substituted sample exhibited a giant coercive field of 1.2 T for the Hall conductivity along with a significant anomalous Hall conductivity of $\sim 500\ \Omega^{-1}cm^{-1}$. Most importantly, the nature of the anomalous Hall conductivity of $Co_3Sn_2S_2$, originating from the topological effects in the band structure, remained intrinsic upon the Ni substitution, further supported by the first-principle calculations. Hence, the topological effects including the Weyl characteristics of $Co_3Sn_2S_2$ were found to be robust against the Ni substitution.

## ASSOCIATED CONTENT

**Supporting Information**.
Compositional analysis by SEM-EDS and wet chemical method; Laue diffraction patterns; anomalous Hall resistivity plots; $\sigma_{xy}$ vs $\sigma_{xx}$ curves; calculated magnetic moment and anomalous Hall conductivity are presented in the supplementary information. "This material is available free of charge via the Internet at http://pubs.acs.org."

## AUTHOR INFORMATION

Corresponding Author
* Nitesh.Kumar@cpfs.mpg.de
* Claudia.Felser@cpfs.mpg.de

ORCID ID

G. S. Thakur: 0000-0002-1362-2357

Author Contributions

The manuscript was written through the contributions of all the authors. All authors have given approval to the final version of the manuscript.

Notes
The authors declare no competing financial interest.

## ACKNOWLEDGMENT

This study was financially supported by the European Research Council Advanced Grant No. 742068 "TOPMAT".

# Table of Contents

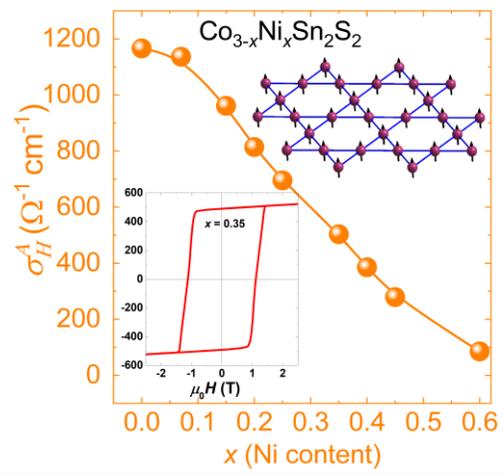